\documentclass[aps,prb,twocolumn,showpacs]{revtex4-1}
\usepackage{amsmath,amssymb,bm,graphicx}
\begin{document}

\title{Optical Aharonov-Bohm Effect on Wigner Molecules
in Type-II Semiconductor Quantum Dots}

\author{Rin Okuyama}
\email{rokuyama@rk.phys.keio.ac.jp}
\author{Mikio Eto}
\author{Hiroyuki Hyuga}
\affiliation{Faculty of Science and Technology,
Keio University,
3-14-1 Hiyoshi, Kohoku-ku, Yokohama 223-8522, Japan}

\date{\today}

\begin{abstract}
We theoretically examine the magnetoluminescence from a
trion and a biexciton in a type-II semiconductor quantum dot,
in which holes are confined inside the quantum dot and electrons
are in a ring-shaped region surrounding the quantum dot.
First, we show that two electrons in the trion and biexciton are
strongly correlated to each other, forming a Wigner molecule:
Since the relative motion of electrons is frozen,
they behave as a composite particle whose mass and charge are twice
those of a single electron. As a result, the energy of the trion and
biexciton oscillates as a function of magnetic field with half
the period of the single-electron Aharonov-Bohm oscillation.
Next, we evaluate the photoluminescence.
Both the peak position and peak height change
discontinuously at the transition of the many-body ground state,
implying a possible observation of the Wigner molecule by
the optical experiment.
\end{abstract}


\maketitle

\section{INTRODUCTION \label{sec:introduction}}

The electron-electron interaction in small rings is an important issue
in mesoscopic physics.
  \cite{Viefers20041}
This is exemplified by unsolved problems in the persistent current
in normal metallic rings. When a magnetic flux penetrates the rings,
the persistent current is induced by the Aharonov-Bohm (AB) effect
in the thermal equilibrium state.
  \cite{Gunther19691391,
        Butiker1983365}
The current observed in experiments
  \cite{PhysRevLett.64.2074,
        PhysRevLett.89.206803,
        PhysRevLett.86.1594,
        PhysRevLett.102.136802,
        Bleszynski-Jayich09102009}
is larger by at least 2 orders of magnitude than the theoretical
prediction for noninteracting electrons.
  \cite{PhysRevLett.62.587,
        0295-5075-8-5-014,
        PhysRevLett.66.88,
        PhysRevB.47.15449}
Besides, the low-flux response is always diamagnetic in experimental
results, whereas it is either diamagnetic or paramagnetic in
theory. To resolve the discrepancies, the theory should take into account
the electron-electron interaction.
  \cite{PhysRevLett.65.381}
The attractive interaction as well as the repulsive one
may play a role in explaining the diamagnetic response
  \cite{0295-5075-13-8-011}
and the magnitude of the persistent current.
  \cite{PhysRevLett.101.057001}

The electron-hole interaction has also been examined in small rings
fabricated on semiconductors. The AB effect on an exciton,
consisting of an electron and a hole,
was theoretically
studied when the electron and hole are confined in small rings of
different size.
   \cite{JETPLett.68.8.669,
         PhysRevB.64.155324,
         PhysRevB.66.081309}
The AB effect becomes different in two situations,
where the electron and hole move almost independently,
or they tightly form an exciton.
The AB effect on an exciton
was observed through photoluminescence
experiments, using quantum rings patterned on InGaAs/GaAs heterostructure
   \cite{PhysRevLett.90.186801}
and InAs/InP quantum tubes.
   \cite{JJAP.46.L440}
This is called the optical AB effect.

\begin{figure}
  \includegraphics[width=7cm]{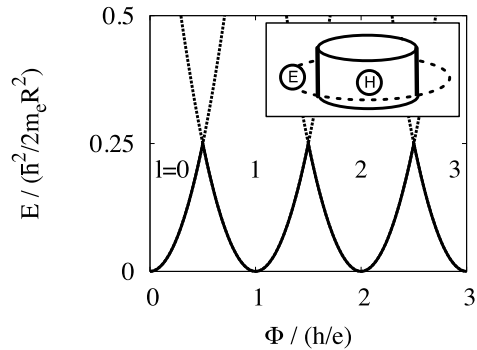}
  \caption{Energy levels of an electron confined
    in a one-dimensional ring of radius $R$,
    as a function of perpendicular magnetic field $B$.
    $\Phi= \pi R^2 B$ is the magnetic flux penetrating
    the ring. The orbital angular momentum $l$ is indicated
    for each energy level.
    Inset: A schematic drawing of the type-II semiconductor
    quantum dot, in which holes are confined inside the quantum
    dot and electrons are in a ring-shaped region surrounding the dot.
    \label{fig:energy_1d}}
\end{figure}

We focus on the type-II semiconductor
quantum dots, such as InP/GaAs, ZnTe/ZnSe, and Ge/Si.
   \cite{PhysRevLett.92.126402,
         PhysRevB.76.035342,
         PhysRevLett.100.136405,
         PhysRevB.77.241302,
         PhysRevB.82.073306}
In these systems, holes are confined inside a quantum dot and electrons
are in a ring-shaped region surrounding the quantum dot, as depicted
in the inset of Fig.\ \ref{fig:energy_1d} (the roles of electron and hole are
exchanged in the case of InP/GaAs). Since the motion of holes
is almost frozen due to the strong confinement, the AB effect on the
electrons can be detected by the photoluminescence.
For an exciton, we can adopt a simple model in which
an electron is confined in a one-dimensional ring with
a perpendicular magnetic field $B$. The Hamiltonian is given by
  \begin{equation}
    H = \frac{\hbar^2}{2 m_\text{e} R^2}
      \left( \hat L - \frac{\Phi}{h/e} \right)^2,
  \label{eq:Hamiltonian0}
  \end{equation}
where $R$ is the radius of the ring,
$m_\text{e}$ is the effective mass of the electron, and
$\hat L$ is the angular momentum operator.
$\Phi = \pi R^2 B$ is the magnetic flux penetrating the ring.
The energy levels are shown in Fig.\ \ref{fig:energy_1d},
as a function of magnetic flux $\Phi$. The quantum number of
the angular momentum $\hat L$, $l$, is indicated for the
respective levels. With an increase in $\Phi$, the ground state
is changed from $l=0$ to $l=1$ at $\Phi=0.5 (h/e)$,
from $l=1$ to $l=2$ at $\Phi=1.5 (h/e)$, and so on.
As a result, the energy of the ground state oscillates as a
function of $\Phi$ with the period of $h/e$.
The magnetic-field dependence of the photoluminescence peak from
an exciton is explained well by this simple model.
   \cite{PhysRevB.66.081309,
         PhysRevLett.92.126402,
         PhysRevB.76.035342,
         PhysRevLett.100.136405,
         PhysRevB.77.241302,
         PhysRevB.82.073306}

In the present paper, we theoretically examine the photoluminescence
from a trion (two electrons and a hole) and a biexciton (two electrons
and two holes) in type-II semiconductor quantum dots,
in order to elucidate the correlation effect between electrons
in a small ring. In our model, the holes are strongly confined in a
harmonic potential, and thus their motion is almost frozen.
This is the experimental situation of Ge/Si quantum dots.
   \cite{PhysRevB.77.241302,
         PhysRevB.82.073306}
The electrons are in a quasi-one-dimensional ring-shaped potential,
$V_\text{e}(r)$, shown in Fig.\ \ref{fig:V}.
First, we calculate the many-body states of a few electrons confined
in $V_\text{e}(r)$ and show the formation of Wigner molecules owing to
the strong correlation effect.
  \cite{Bolton1993139,
        PhysRevB.53.10871,
        PhysRevB.55.13707,
        PhysRevLett.85.1726,
        Manninen.2001}
In the Wigner molecules of $N$ electrons,
the electrons behave as a single
particle whose mass and charge are $N$ times of those of an electron.
In consequence, the energy of the ground state oscillates
with $\Phi$ by the period of $h/(Ne)$, the so-called fractional AB effect.
  \cite{0953-8984-3-18-014,
        PhysRevB.45.11795,
        PhysRevB.49.16234,
        0295-5075-36-7-533}
The formation of Wigner molecules is also seen for electrons in
trions and biexcitons.
Next, we examine the photoluminescence from the electron-hole
complexes. We observe that the peak position and intensity of
the photoluminescence, as a function of $\Phi$,
change discontinuously at the transition of the ground state.
This implies a possible observation of the Wigner molecules by
the optical experiment. We hope that our prediction will
motivate the experimental study of trions and biexcitons in
type-II quantum dots although there have not been such
experiments until now.

We should make a comment on the cylindrical symmetry in our model
as well as in the Hamiltonian in Eq.\ (\ref{eq:Hamiltonian0}).
As mentioned above, the peak position of the luminescence from an
exciton can be
explained using the Hamiltonian in Eq.\ (\ref{eq:Hamiltonian0}),
but the peak intensity cannot. For the optical recombination,
the total angular momentum $L$ of an electron and a hole must be zero
because the final state is the vacuum with no electron or hole.
Since the angular momentum of the hole is assumed to be zero
in the ground state,
the recombination is possible only at $\Phi \le 0.5(h/e)$
where the angular momentum of the electron is $l=0$.
   \cite{JETPLett.68.8.669}
This contradicts the experimental results which observed
a finite intensity at $\Phi > 0.5(h/e)$.
  \cite{PhysRevLett.92.126402,
        PhysRevLett.100.136405,
        PhysRevB.82.073306}
This discrepancy could be resolved if the disorder of the system and
finite temperature were taken into account.
   \cite{PhysRevB.70.155318,
         PhysRevB.76.195326,
         PhysRevB.78.075322}
In our study we do not consider the disorder
effect, which would modify our calculated results about the peak
intensity of photoluminescence. On the other hand,
our results about the peak position should not be changed
qualitatively by the disorder.
In particular, the discontinuous
change of the peak position, which directly reflects the Wigner
molecularization, can be experimentally observed
although it is smeared to some extent.
The discontinuous change of the peak intensity could be
observed if well-shaped samples were fabricated.

The present paper is organized as follows.
In Sec.\ \ref{sec:model}, we present our model and calculation
method. We adopt the exact diagonalization method for the many-body
states of electrons and holes.
In Sec.\ \ref{sec:electron}, we begin with the many-body
states of a few electrons confined in the ring-shaped region.
No holes are considered. We show the formation of Wigner molecules,
reflecting the strong correlation effect.
In Sec.\ \ref{sec:electron-hole}, we calculate the many-body states
of trions and biexcitons, in which two electrons form a Wigner molecule
despite the presence of the holes. We evaluate the photoluminescence,
using the many-body states obtained by the exact diagonalization
method. Finally the conclusions are given in Sec.\ \ref{sec:conclusions}.

\section{MODEL AND CALCULATION METHOD \label{sec:model}}

\subsection{Effective-mass Hamiltonian}

We consider a type-II semiconductor quantum dot of cylindrical
symmetry in the $xy$ plane. Holes are localized inside a
quantum dot of disk shape, whereas electrons are confined in a
ring-shaped region surrounding the dot. A magnetic field is applied
perpendicularly to the quantum dot.

We adopt the effective-mass approximation,
assuming that the radius of the quantum dot, $R \gtrsim 10~$nm, is
much larger than the lattice constant $a$.
  \cite{PhysRevLett.92.126402,
        PhysRevB.76.035342,
        PhysRevLett.100.136405,
        PhysRevB.77.241302,
        PhysRevB.82.073306}
For the holes, only the heavy-hole band (total angular momentum
$j=3/2$, $j_z=\pm3/2$) is considered because
the confinement in the $z$ direction splits it from
the light-hole band ($j=3/2$, $j_z=\pm1/2$).
  \footnote{
    See, e.g., Chap. 4 of H. Haug and S. W. Koch,
    {\it Quantum Theory of
    the Optical and Electronic Properties of Semiconductors},
    Fifth Edition
    (World Scientific, Singapore, 2009),
    or Chap. 9 of
    P. Y. Yu and M. Cardona,
    {\it Fundamentals of Semiconductors}, Fourth Edition
    (Springer, Berlin, 2010).
  }
The wave functions of the electron and hole are written as
  \begin{eqnarray}
    \Psi_{\text{e},\pm} &=& \psi_{\text{e}}(\bm{r})~
    u_{\text{c}}(\bm{r}) \chi_{\pm},
    \label{eq:wavef-e}
    \\
    \Psi_{\text{h},\pm} &=& \psi_{\text{h}}(\bm{r})~
    u^*_{\text{v},\mp} (\bm{r}) \chi_{\mp},
  \label{eq:wavef-h}
  \end{eqnarray}
respectively, where $\chi_{\pm}$ indicates the spin-up
($s_z=1/2$) or -down ($s_z=-1/2$).
$\psi_\text{e}(\bm{r})$ and $\psi_\text{h}(\bm{r})$
are envelope functions for electrons and holes, whereas
$u_{\text{c}}(\bm{r})$ and $u_{\text{v},\pm} (\bm{r})$
are the Bloch function of the conduction band ($s$ wave) and
valence band (orbital angular momentum $l_z=\pm 1$) at the
$\Gamma$ point, respectively.
\footnote{
$\bm{r}$ in the Bloch functions (and integral of $\int^{\text{(3D)}}$)
is a three-dimensional vector, describing the scale of
lattice constant. In the other places, $\bm{r}$ is
a two-dimensional vector for the larger scale.
}

\begin{figure}
  \includegraphics[width=7cm]{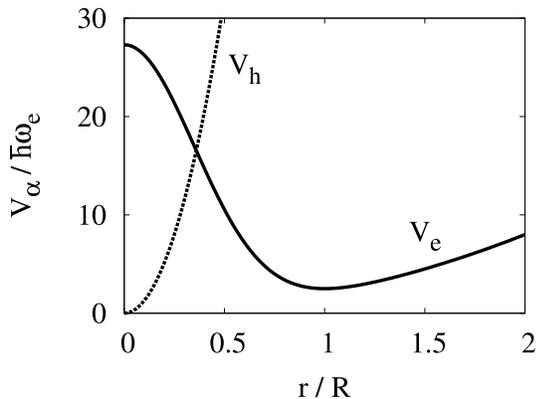}
  \caption{The confinement potential for
  electrons, $V_\text{e}(r)$, and that for
  holes, $V_\text{h}(r)$, as a function of
  the radial coordinate $r$.
  $R$ is the radius at which $V_\text{e}(r)$ takes a minimum.
  \label{fig:V}}
\end{figure}

The confinement potential for the holes is given by
a harmonic potential
  \begin{equation}
    V_\text{h} (r) = \frac{1}{2} m_\text{h} \omega_\text{h}^2 r^2,
  \label{eq:Vh}
  \end{equation}
while that for the electrons is
  \begin{equation}
    V_\text{e} (r) = \frac{1}{2} m_\text{e} \omega_\text{e}^2 r^2
      + V_0 \exp(-\alpha r^2),
  \label{eq:Ve}
  \end{equation}
where $m_\text{h}$ and $m_\text{e}$ are the effective masses
of holes and electrons, respectively.
The parameters $\omega_\text{h}, \omega_\text{e}, V_0,$ and $\alpha$
are chosen so that the radius of the electron confinement
$R$ at which $V_\text{e}(r)$ takes a minimum
is eight times larger than the radius of hole confinement
$\sqrt{\hbar / m_\text{h} \omega_\text{h}}$.
This choice confirms the strong confinement of holes in a quantum
dot, in accordance with the experimental situation.
   \cite{PhysRevB.77.241302,
         PhysRevB.82.073306}
The confinement potentials $V_\text{h} (r)$ and $V_\text{e}(r)$
are depicted in Fig.\ \ref{fig:V}.

The envelope functions $\psi_{\text{e}}(\bm{r})$ and
$\psi_{\text{h}}(\bm{r})$ are determined from the
effective-mass Hamiltonian. For $N_\text{e}$ electrons and
$N_\text{h}$ holes, the Hamiltonian is given by
\begin{widetext}
  \begin{eqnarray}
    H &=& H_\text{e} + H_\text{h} + H_\text{e-h},
    \label{eq:H} \\
    H_\text{e} &=& \sum_{j=1}^{N_\text{e}}
      \left\{ \frac{1}{2 m_\text{e}}
        \Bigl[ - i \hbar \frac{\partial}{\partial \bm{r}_{\text{e},j}}
               + e \bm{A}(\bm{r}_{\text{e},j}) \Bigr]^2
               + V_\text{e}(\bm{r}_{\text{e},j}) \right\}
      + \sum_{1 \leq j < k \leq N_\text{e}} \frac{e^2}{4 \pi \epsilon
          |\bm{r}_{\text{e},j} - \bm{r}_{\text{e},k}|},
    \label{eq:He} \\
    H_\text{h} &=& \sum_{j=1}^{N_\text{h}}
      \left\{ \frac{1}{2 m_\text{h}}
        \Bigl[ - i \hbar \frac{\partial}{\partial \bm{r}_{\text{h},j}}
               - e \bm{A}(\bm{r}_{\text{h},j}) \Bigr]^2
               + V_\text{h}(\bm{r}_{\text{h},j}) \right\}
      + \sum_{1 \leq j < k \leq N_\text{h}} \frac{e^2}{4 \pi \epsilon
          |\bm{r}_{\text{h},j} - \bm{r}_{\text{h},k}|},
    \label{eq:Hh} \\
    H_\text{e-h} &=&
        \sum_{j=1}^{N_\text{e}} \sum_{k=1}^{N_\text{h}}
          \frac{-e^2}{4 \pi \epsilon
          |\bm{r}_{\text{e},j} - \bm{r}_{\text{h},k}|}.
  \end{eqnarray}
The magnetic field $B$ is applied in the $-z$ direction:
$\nabla \times \bm{A}(\bm{r}) = - B \mathbf{e}_z$.
$H_\text{e}$ $(H_\text{h})$ is the Hamiltonian
for interacting electrons (holes), whereas
$H_\text{e-h}$ describes the electron-hole interaction.
We assume that the dielectric constant $\epsilon$ is identical
for electrons and holes.
The spin Zeeman effect is neglected.

In $H_\text{e-h}$, the exchange-type interaction between
an electron and a hole is disregarded for the following reason.
For the wave functions in Eqs.\ (\ref{eq:wavef-e}) and
(\ref{eq:wavef-h}), the matrix element of
\[
\langle \Psi_{\text{e},\sigma}^{\prime}
\Psi_{\text{h},-\sigma}^{\prime} |
H_\text{e-h} | \Psi_{\text{h},-\sigma} \Psi_{\text{e},\sigma}
\rangle
=\int^\text{(3D)} \mathrm{d}\bm{r}_1 \mathrm{d}\bm{r}_2
      \Psi_{\text{e},\sigma}^{\prime *}(\bm{r}_1)
      \Psi_{\text{h},-\sigma}^{\prime *}(\bm{r}_2)
      \frac{-e^2}{4\pi\epsilon} \frac{1}{|\bm{r}_1 - \bm{r}_2|}
      \Psi_{\text{h},-\sigma}(\bm{r}_1)
      \Psi_{\text{e},\sigma}(\bm{r}_2)
\]
for $\sigma=\pm$,\cite{Note2} involves the integral of
$u_{\text{c}}^{*}(\bm{r}_1) u_{\text{v},\sigma}(\bm{r}_1)$,
which oscillates with the period of the lattice constant $a$.
In consequence, the matrix element is
smaller by the order of $(a/R)^2 \sim 10^{-4}$ than
the exchange interaction between two electrons or that between
two holes. Therefore we only consider the matrix elements
of $\langle \Psi_{\text{e},\sigma}^{\prime}
\Psi_{\text{h},\sigma^{\prime}}^{\prime} |
H_\text{e-h} | \Psi_{\text{e},\sigma} \Psi_{\text{h},\sigma^{\prime}}
\rangle$ for $H_\text{e-h}$:
\[
\langle \Psi_{\text{e},\sigma}^{\prime}
\Psi_{\text{h},\sigma^{\prime}}^{\prime} |
H_\text{e-h} | \Psi_{\text{e},\sigma} \Psi_{\text{h},\sigma^{\prime}}
\rangle
=\int \mathrm{d}\bm{r}_1 \mathrm{d}\bm{r}_2
      \psi_{\text{e}}^{\prime *}(\bm{r}_1)
      \psi_{\text{h}}^{\prime *}(\bm{r}_2)
      \frac{-e^2}{4\pi\epsilon} \frac{1}{|\bm{r}_1 - \bm{r}_2|}
      \psi_{\text{e}}(\bm{r}_1)
      \psi_{\text{h}}(\bm{r}_2)
\]
\end{widetext}
after the integrations of $|u_\text{c}(\bm{r}_1)|^2$ and
$|u_\text{v}(\bm{r}_2)|^2$ over the unit cell of the lattice.

The strength of the magnetic field is measured by
the flux penetrating the ring of radius $R$, $\Phi = \pi R^2 B$.
The ratio of the strength of the Coulomb potential to the kinetic energy
is characterized by the parameter of $R / a_\text{\tiny B}$,
where $a_\text{\tiny B} = 4 \pi \epsilon \hbar^2 / (m_\text{e} e^2)$
is the effective Bohr radius for electrons.
We assume that $R / a_\text{\tiny B} \gtrsim 1$, considering
the experimental situations.
  \cite{PhysRevLett.92.126402,
        PhysRevB.76.035342,
        PhysRevLett.100.136405,
        PhysRevB.77.241302,
        PhysRevB.82.073306}

\subsection{Exact diagonalization method}

The many-body states of electrons and holes are calculated by
the exact diagonalization method, taking full account of the
Coulomb interaction.
As a basis set, we adopt the eigenfunctions of the one-body
part of $H_{\text{e}}$ in Eq.\ (\ref{eq:He}) for electrons
and those of $H_{\text{h}}$ in Eq.\ (\ref{eq:Hh}) for holes.
  \footnote{
  The basis functions are analytically obtained for the holes.
  Those for electrons are expressed as a linear combination
  of the eigenfunctions of $H_{\text{e},0} =
      (\bm{p}+e \bm{A})^2/(2 m_\text{e})
               + m_\text{e} \omega_\text{e}^2 r^2/2$,
  the coefficients of which are numerically determined.
  }
They are denoted by $\psi_{\text{e},l,n}(\bm{r})$ and
$\psi_{\text{h},l,n}(\bm{r})$, respectively, with
quantum number of orbital angular momentum
($l=0,\pm 1, \pm 2, \ldots$) and that of radial motion
($n = 1,2,3,\ldots$).

Note that the effective-mass Hamiltonian in Eq.\ (\ref{eq:H})
has an axial symmetry in space
and that the electron spins are decoupled from the hole spins.
Therefore, the total orbital angular momentum $L$,
total electron spin $(S_\text{e}, S_{\text{e},z})$,
and total hole spin $(S_\text{h}, S_{\text{h},z})$
are good quantum numbers.
The energy eigenvalues do not depend on
$S_{\text{e},z}$ and $S_{\text{h},z}$.
Hence we diagonalize the Hamiltonian in the subspace with
given values of $L$, $S_{\text{e},z}$, and $S_{\text{h},z}$,
the dimensions of which are less than $10^4$.
\footnote{
    For example, the dimension is 4,339 for the states of
    three electrons with $L=0$ and $S_{\text{e},z}=1/2$,
    and 21,025 for the states of biexciton with $L=0$,
    $S_{\text{e},z}=0$, and $S_{\text{h},z}=0$.
}
The truncation of the Hamiltonian matrix leads to
an inaccuracy of 0.1\% for the total energy and
of 1\% for the intensity of the photoluminescence.

\subsection{Photoluminescence \label{subsec:spectrum}}

We evaluate the photoluminescence from a trion and
a biexciton, using the many-body states obtained by the
exact diagonalization method. We assume that the initial
state is the ground state of the trion or biexciton.

When an electron with spin-up ($\Psi_{\text{e},+}$) recombines
with a hole with spin-down ($\Psi_{\text{h},+}$), a 
right-circular photon is emitted. Similarly, when
an electron with spin-down ($\Psi_{\text{e},-}$) recombines
with a hole with spin-up ($\Psi_{\text{h},-}$), a
left-circular photon is emitted. The recombination
rate is evaluated by Fermi's golden rule with the
dipole approximation.
  \cite{FizTekhPoluprov.3.1042,
        UspFizNauk.136.459}
The interband dipole-moment operator is given by
  \begin{equation}
    \hat d = \sum_{l,n_1,n_2} \sum_{\sigma=\pm} d_{l,n_1,n_2}
      \hat e_{l,n_1,\sigma} \hat h_{-l,n_2,\sigma} + \text{h.c.},
  \end{equation}
where $\hat e_{l,n,\sigma}$ [$\hat h_{l,n,\sigma}$] is an
annihilation operator of an electron [a hole] in the state of
$\psi_{\text{e},l,n} \chi_{\sigma}$
[$\psi_{\text{h},l,n} \chi_{-\sigma}$] and
  \begin{eqnarray}
    d_{l,n_1,n_2} &=& d_\text{vc} \int \mathrm{d}\bm{r}~
      \psi_{\text{e},l,n_1}(\bm{r})~\psi_{\text{h},-l,n_2}(\bm{r}),
    \label{eq:d-vc0} \\
    d_\text{vc} &=& \left| \int_{\text{unit cell}}^\text{(3D)}
      \mathrm{d}\bm{r} ~ u^*_{\text{v},\sigma}(\bm{r}) ~
      (-e \bm{r}) ~ u_\text{c}(\bm{r}) \right|.
    \label{eq:d-vc}
  \end{eqnarray}
$d_\text{vc}$ is independent of $\sigma=\pm$.
The transition rate from the initial state
$| \text{init} \rangle$ with energy $E_\text{init}$ to
the final state $|\text{fin}\rangle$ with energy $E_\text{fin}$
is written as
\cite{BallentineQM}
  \begin{eqnarray}
    I &=& \frac{4}{3} \frac{E^3}{4\pi\epsilon_0 \hbar^4c^3}
      | \langle \text{fin} | \hat d | \text{init} \rangle |^2,
   \label{eq:intensity-formula}
  \end{eqnarray}
which is accompanied by the photon emission of energy
$E=E_\text{init}-E_\text{fin}$ ($\simeq E_\text{gap}$, band gap).
Note that the total angular momentum $L$
and the total spin $S_{\text{e},z} + S_{\text{h},z}$
should be conserved during the transition, as seen in
Eq.\ (\ref{eq:d-vc0}).
The intensity of the photoluminescence is evaluated by $I$,
with $E$ being replaced by $E_\text{gap}$.

\section{Wigner Molecules of FEW ELECTRONS \label{sec:electron}}

We begin with the many-body states of a few electrons confined
in a ring-shaped potential $V_\text{e} (r)$ in Eq.\
(\ref{eq:Ve}), to illustrate the Wigner molecularization.
The number of electrons is $N_\text{e}=1$ to 3.
No holes are assumed in this section.
The many-body states and energies are
obtained using the exact diagonalization method for the
Hamiltonian $H_\text{e}$ in Eq.\ (\ref{eq:He}).

\begin{figure}
  \includegraphics[width=7cm]{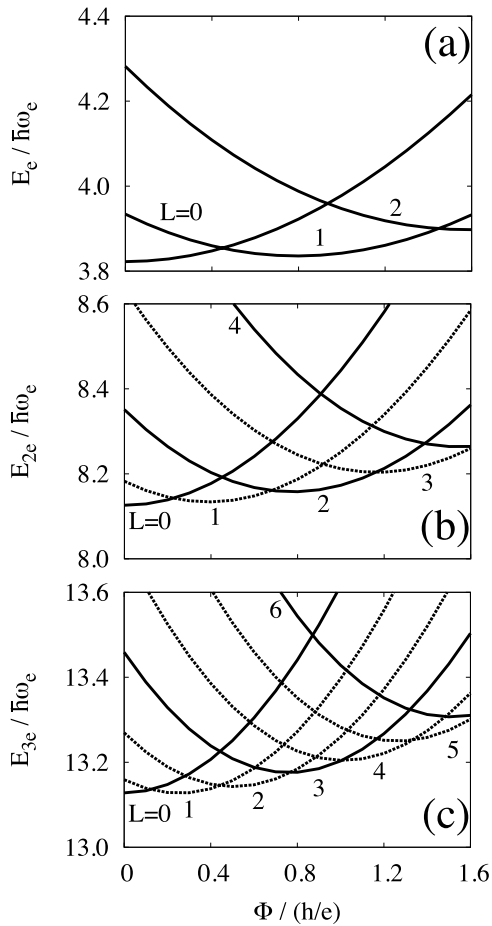}
  \caption{Low-lying energies for (a) one, (b) two, and (c)
    three electrons confined in a ring-shaped potential,
    $V_\text{e}(r)$, as a function of the magnetic flux $\Phi$.
    The radius $R$ at which $V_\text{e}(r)$ takes a minimum
    equals the effective Bohr radius $a_\text{\tiny B}$.
    $\Phi= \pi R^2 B$.
    Solid and broken lines indicate spin-singlet and
    -triplet states, respectively, in (b), and
    spin-quartet and -doublet states in (c).
  \label{fig:energy_few_e}}
\end{figure}

The calculated results of the energies are shown in
Fig.\ \ref{fig:energy_few_e} as a function of magnetic
flux $\Phi$, for (a) $N_\text{e}=1$, (b) 2, and (c) 3.
$R/a_\text{B} = 1$.
The total angular momentum $L$ is indicated for
respective states.
For one electron, the transition of the ground state
takes place at $\Phi \simeq 0.5$ and $1.5$ in units of $h/e$.
The angular momentum increases by one at each transition.
The $\Phi$ dependence of the energies is similar to
that in Fig.\ \ref{fig:energy_1d} for an electron in a
one-dimensional ring
although the diamagnetic shift of the energies is seen
in Fig.\ \ref{fig:energy_few_e}(a).
Therefore, the energy of the ground state oscillates
with the period of approximately $h/e$, reflecting the
one-dimensional motion along the ring.

Note that the state of $L$ in Fig.\ \ref{fig:energy_few_e}(a)
corresponds to $\psi_{\text{e},L,1}$ introduced in Sec.\ II.B.
The energies of the excited states in the radial motion,
$\psi_{\text{e},L,n}$ $(n \ge 2)$, are larger by more than
2$\hbar \omega_\text{e}$ than those of the lowest states.
In the low-lying states of two and three electrons, shown
in Figs.\ \ref{fig:energy_few_e}(b) and \ref{fig:energy_few_e}(c),
the weight of
$\psi_{\text{e},L,n}$ $(n \ge 2)$ is of the order of $10^{-3}$.
Therefore, electrons possess a one-dimensional nature in our model.

In Figs.\ \ref{fig:energy_few_e}(b) and \ref{fig:energy_few_e}(c),
we observe the
change of the ground state for two and three
electrons. If the diamagnetic shift is disregarded, the
energy of the ground state oscillates quasiperiodically
with the period of $h/(2e)$ for two electrons and
$h/(3e)$ for three electrons. This implies the formation
of Wigner molecules as explained below.

\begin{figure}
  \includegraphics[width=7cm]{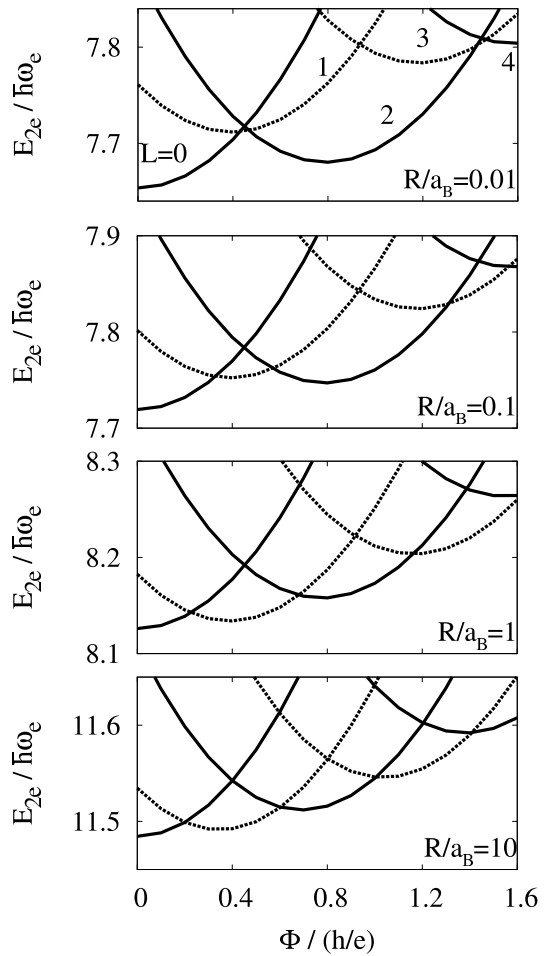}
  \caption{Low-lying energies for two electrons
    confined in a ring-shaped potential, $V_\text{e}(r)$,
    as a function of the magnetic flux $\Phi$.
    $R/a_\text{\tiny B} = 0.01$, 0.1, 1, and 10, where
    $R$ is the radius at which $V_\text{e}(r)$ takes a minimum
    and $a_\text{\tiny B}$ is the effective Bohr radius.
    $\Phi= \pi R^2 B$.
    Solid and broken lines indicate spin-singlet and
    -triplet states, respectively.
  \label{fig:energy_2e}}
\end{figure}

In order to elucidate the correlation effect,
we examine many-body states for two electrons with changing
$R/a_\text{B}$.
Figure \ref{fig:energy_2e} shows low-lying energies
in the case of $R/a_\text{B} = 0.01$, $0.1$, $1$, and $10$.
When the Coulomb interaction is very weak
($R/a_\text{B} = 0.01$),
two electrons occupy the lowest orbital shown
in Fig.\ \ref{fig:energy_few_e}(a) in the ground state.
Consequently, the total angular momentum is always even
and the total spin is a singlet.
As the strength of the Coulomb interaction
increases with $R/a_\text{B}$,
the exchange interaction lowers the energy of the
spin-triplet states with $L=1$ and $3$.
For $R/a_\text{B} \gtrsim 1$,
spin-singlet and -triplet states appear alternatively
as $\Phi$ increases by approximately $h/(2e)$.
Thus the fractional period of the energy oscillation is
ascribable to the strong correlation effect.
Note that the period of the energy oscillation
in the case of $R/a_\text{B} = 10$ is slightly
shorter than that of $R/a_\text{B} = 1$.
This is because the Coulomb repulsion between electrons
increases the expectation value of the electron radius.

\begin{figure}
  \includegraphics[width=7cm]{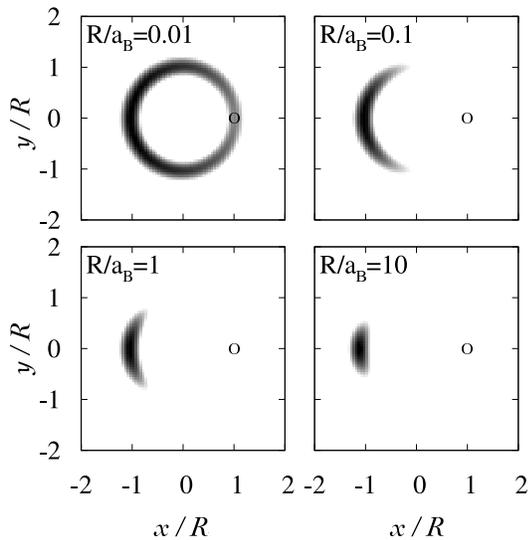}
  \caption{Gray scale plots of the two-body density
    for two electrons confined in a ring-shaped potential,
    $V_\text{e}(r)$. The magnetic field is $B=0$.
    $R/a_\text{\tiny B} = 0.01$, 0.1, 1, and 10, where
    $R$ is the radius at which $V_\text{e}(r)$ takes a minimum
    and $a_\text{\tiny B}$ is the effective Bohr radius.
    One electron is fixed at the point indicated by an open circle.
  \label{fig:rho_2e}}
\end{figure}

To examine the correlation effect further,
we calculate the two-body density
  \begin{equation} \label{eq:rho}
    \rho(\bm{r}|\bm{r}_0) =
    \frac12 \sum_{\sigma,\sigma_0}
    \left\langle
    \hat\psi^\dagger_{\text{e},\sigma} (\bm{r})
    \hat\psi^\dagger_{\text{e},\sigma_0} (\bm{r}_0)
    \hat\psi_{\text{e},\sigma_0} (\bm{r}_0)
    \hat\psi_{\text{e},\sigma} (\bm{r})
    \right\rangle,
  \end{equation}
where $\hat\psi_{\text{e},\sigma} (\bm{r})$ and
$\hat\psi^\dagger_{\text{e},\sigma} (\bm{r})$ are the
field operators of electron with spin $\sigma$.
Figure \ref{fig:rho_2e} shows $\rho(\bm{r}|\bm{r}_0)$
for two electrons in the ground state at the magnetic
flux $\Phi=0$. $\bm{r}_0$ is fixed at the position
indicated by an open circle.
$R / a_\text{B}$ is changed in the same way as in Fig.\
\ref{fig:energy_2e}. For $R / a_\text{B} \gtrsim 1$,
two electrons maximize their distance by
being localized at the other side of each other in the ring.
This clearly indicates the formation of the Wigner molecule.
Since the relative motion is frozen, the two electrons
behave as a composite particle
whose mass and charge are twice those of an electron.
In consequence the ground-state energy
oscillates with $\Phi$ by the period of $h/(2e)$.

\begin{figure}
  \includegraphics[width=7cm]{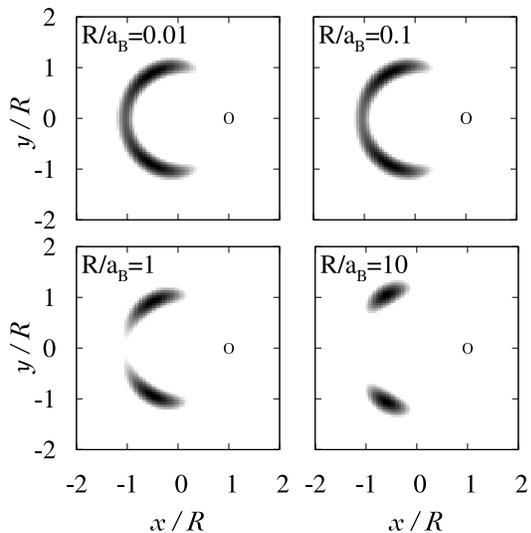}
  \caption{Gray scale plots of the two-body density
    for three electrons confined in a ring-shaped potential,
    $V_\text{e}(r)$. The magnetic field is $B=0$.
    $R/a_\text{\tiny B} = 0.01$, 0.1, 1, and 10, where
    $R$ is the radius at which $V_\text{e}(r)$ takes a minimum
    and $a_\text{\tiny B}$ is the effective Bohr radius.
    One electron is fixed at the point indicated by an open circle.
  \label{fig:rho_3e}}
\end{figure}

For three electrons, a similar formation of
the Wigner molecule is observed for $R / a_\text{B} \gtrsim 1$.
Figure \ref{fig:rho_3e} shows
the two-body density for three electrons in the ground state
at $\Phi=0$. The electrons are localized
around apices of an equilateral triangle in the ring.
The molecularization explains the energy oscillation
with the period of $h/(3e)$ in Fig.\ \ref{fig:energy_few_e}(c).

We make a comment on the total spin $S$ of the ground
state. The spin $S$ changes with the total
angular momentum $L$ at the transition of the ground
state shown in Fig.\ \ref{fig:energy_few_e}.
For two electrons, as we mentioned above,
$S = 0$ ($S = 1$) when $L$ is an even (odd).
For three electrons,
$S = 3/2$ if $L$ is a multiple of 3
and $S = 1/2$ otherwise.
This was explained by the $N_\text{e}$-fold rotational symmetry
of $N_\text{e}$-electron configuration in the Wigner molecule.
    \cite{PhysRevB.63.205323,
          Koskinen.2002,
          PhysRevB.73.113313}

\section{ELECTRON-HOLE COMPLEXES AND OPTICAL RESPONSE
\label{sec:electron-hole}}

In this section,
we examine the many-body states of electron-hole complexes,
that is, exciton, trion, and biexciton.
We consider the case of $R/a_\text{B} = 1$,
in which two electrons form a Wigner molecule in the cases of
trion and biexciton.
First, the low-lying states are analyzed as a function of
magnetic flux $\Phi$.
Then the photoluminescence is examined from the ground state in
trion and biexciton.

\subsection{Low-lying states}

\begin{figure}
  \includegraphics[width=7cm]{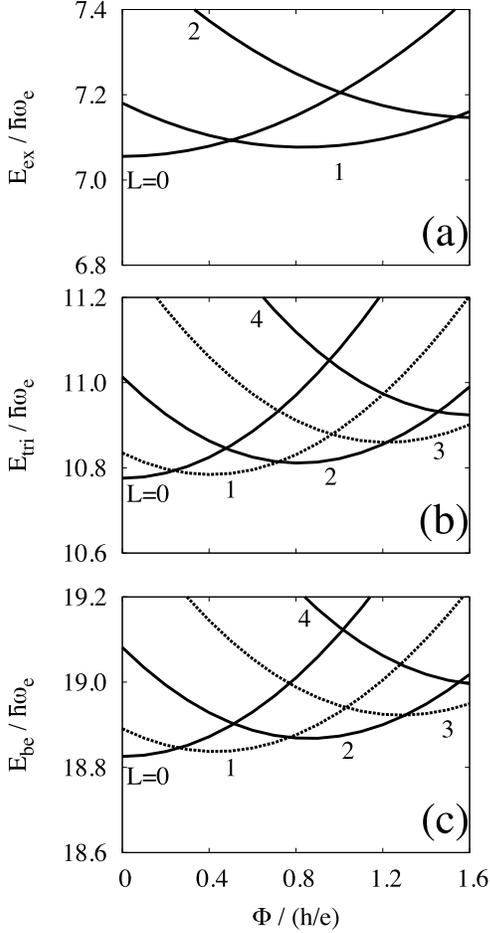}
  \caption{Low-lying energies for (a) exciton, (b) trion, and
    (c) biexciton, as a function of the magnetic flux $\Phi$.
    The radius $R$, at which $V_\text{e}(r)$ takes a minimum,
    equals the effective Bohr radius $a_\text{\tiny B}$.
    $\Phi= \pi R^2 B$. In (b) and (c),
    solid and broken lines indicate the spin states of electrons:
    spin-singlet and -triplet states, respectively.
  \label{fig:energy_eh}}
\end{figure}

Figure \ref{fig:energy_eh} shows the low-lying energies for
(a) exciton, (b) trion, and (c) biexciton, as a function of
magnetic flux $\Phi$.
For an exciton in Fig.\ \ref{fig:energy_eh}(a),
the angular momentum of the ground state changes at
$\Phi \approx 0.5(h/e)$ and $1.5(h/e)$, which is
qualitatively the same as in Fig.\ \ref{fig:energy_few_e}(a)
for an electron confined in $V_\text{e}(r)$. This is because
the hole occupies the lowest state with angular momentum $l=0$
and is insensitive to the magnetic field.

For the trion and biexciton in Figs.\ \ref{fig:energy_eh}(b)
and \ref{fig:energy_eh}(c),
the ground state changes in a similar manner to that for
two electrons confined in $V_\text{e}(r)$ [Fig.\
\ref{fig:energy_few_e}(b)].
A hole or two holes occupy the lowest state with angular momentum
$l=0$, which is hardly influenced by the magnetic field.
Two electrons in a trion and a biexciton form a Wigner molecule,
which is reflected by the energy oscillation with the period
of $h/(2e)$.

Precisely speaking, the magnetic flux $\Phi$ at the transition of
the ground state is slightly shifted to the larger values
for the trion [Fig.\ \ref{fig:energy_eh}(b)] and
biexciton [Fig.\ \ref{fig:energy_eh}(c)],
compared with the two-electron case [Fig.\ \ref{fig:energy_few_e}(b)].
This is because
one hole or two holes inside the quantum dot decrease the
effective radius of electrons. Although the screening
by the holes should weaken the electron-electron interaction,
its effect is invisible: The screening effect on the Wigner
molecule is negligible unless it is so large as to break
the molecule.

\subsection{Photoluminescence}

\begin{figure}
  \includegraphics[width=7cm]{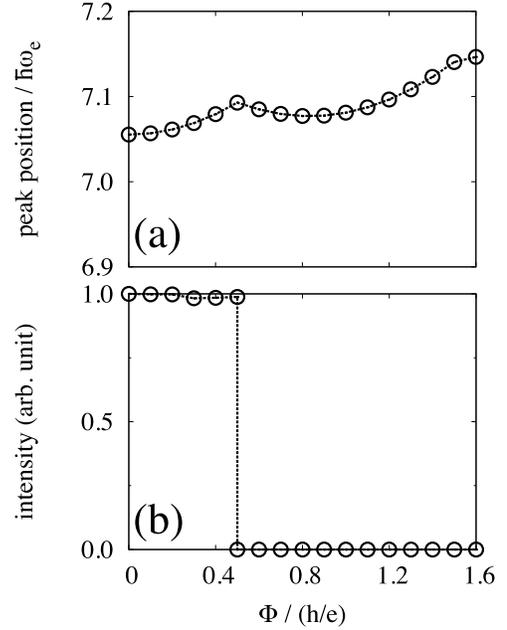}
  \caption{
    (a) The peak position and (b) intensity of the
    photoluminescence from an exciton,
    as a function of the magnetic flux $\Phi$.
    The radius $R$, at which $V_\text{e}(r)$ takes a minimum,
    equals the effective Bohr radius $a_\text{\tiny B}$.
    $\Phi= \pi R^2 B$.
    \label{fig:exciton}}
\end{figure}

Now we discuss the photoluminescence from the
electron-hole complexes.
Figure \ref{fig:exciton} shows the $\Phi$ dependence of
(a) peak position and (b) intensity of the photoluminescence
from an exciton. The peak position
coincides with the ground-state energy shown in Fig.\
\ref{fig:energy_eh} because the final state is the vacuum.
The optical recombination of the exciton is possible
when the angular momentum of the electron is $l=0$ since
that of the hole is always $l=0$. In consequence the
exciton gets dark after the first transition
of the electronic state at $\Phi \simeq 0.5(h/e)$,
as mentioned in Sec.\ I.

In the photoluminescence from a trion, there are two electrons and
a hole in the initial state, and an electron in the final state.
If the ground state of the trion has the angular momentum $L$,
it changes to a one-electron state of $\psi_{\text{e},L,n}$
$(n=1,2,\ldots)$. The transition to $\psi_{\text{e},L,1}$ is
dominant because the weight of $\psi_{\text{e},L,n}$ ($n \ge 2$)
is very small in the ground state. The transition to the higher
states is neglected since the intensity is $10^{-3}$ times as small
as that of the dominant transition.

\begin{figure}
  \includegraphics[width=7cm]{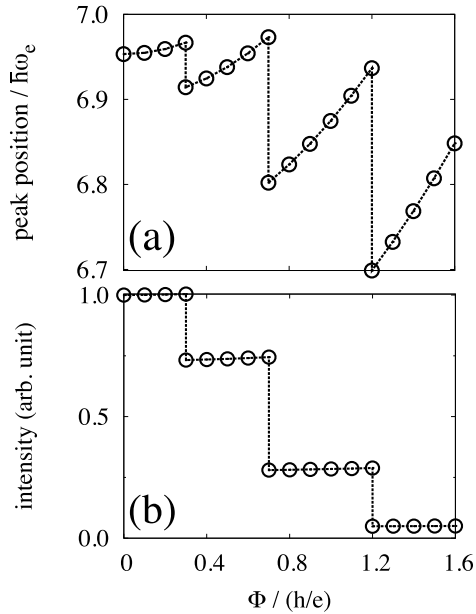}
  \caption{
    (a) The peak position and (b) intensity of the
    photoluminescence from a trion,
    as a function of the magnetic flux $\Phi$.
    The radius $R$, at which $V_\text{e}(r)$ takes a minimum,
    equals the effective Bohr radius $a_\text{\tiny B}$.
    $\Phi= \pi R^2 B$.
    \label{fig:trion}}
\end{figure}

The position and intensity of the dominant peak from a trion
are shown in Fig.\ \ref{fig:trion}, as a function of magnetic
flux $\Phi$.
The peak position increases with an increase in $\Phi$,
and suddenly drops when the angular momentum $L$ is changed
in the ground state of the trion. At the transition of $L$,
the final state is changed. As a result, the energy of the
final state, $E_{\text{e}}$, is discontinuously changed,
whereas the energy of the initial state, $E_{\text{2e}}$, is
continuous as shown in Fig.\ \ref{fig:energy_eh}(b).

As seen in Fig.\ \ref{fig:trion}(b), the intensity of the
photoluminescence from a trion shows a plateau structure
as a function of $\Phi$: While the angular momentum $L$ is
not changed in the ground state, the intensity is almost
constant. At the transition of $L$, the intensity decreases
abruptly. The height of the plateaus indicates a ratio
of 4~:~3~:~1~:~0 approximately.

The simple ratio of the intensity is explained
in the following. The strongly correlated
two-electron states can be approximated by a few electronic
configurations:
\begin{widetext}
  \begin{eqnarray}
    |L=0\rangle &=&
    \left[
      \sqrt{\frac{2}{3}} \hat e^\dagger_{0, +}
                         \hat e^\dagger_{0, -}
      - \sqrt{\frac{1}{6}} \left(
          \hat e^\dagger_{1,+}   \hat e^\dagger_{-1,-}
        - \hat e^\dagger_{1,-} \hat e^\dagger_{-1,+}
      \right)
    \right] |0\rangle,
   \label{eq:HL0}
   \\
    |L=1\rangle &=&
    \frac{1}{\sqrt{2}}
    \left( \hat e^\dagger_{0,+} \hat e^\dagger_{1,-}
    + \hat e^\dagger_{0,-} \hat e^\dagger_{1,+}
    \right) |0\rangle,
   \label{eq:HL1}
   \\
    |L=2\rangle &=&
    \left[
      \sqrt{\frac{2}{3}} \hat e^\dagger_{1, +}
                         \hat e^\dagger_{1, -}
    - \sqrt{\frac{1}{6}} \left(
      \hat e^\dagger_{2,+}   \hat e^\dagger_{0,-}
    - \hat e^\dagger_{2,-} \hat e^\dagger_{0,+}
    \right) \right] |0\rangle,
   \label{eq:HL2}
  \end{eqnarray}
\end{widetext}
where $\hat e^\dagger_{l,\sigma}$ is the creation operator of
an electron in state $\psi_{\text{e},l,1}\chi_{\sigma}$
(see the Appendix).
These expressions are an extension of the Heitler-London
wave function for two electrons in a hydrogen molecule.
  \cite{ZPhys.44.455}
From the wave functions in Eqs.\ (\ref{eq:HL0})--(\ref{eq:HL2}),
we obtain the intensity ratio of 4~:~3~:~1~:~0, as explained in
the Appendix.

\begin{figure}
  \includegraphics[width=7cm]{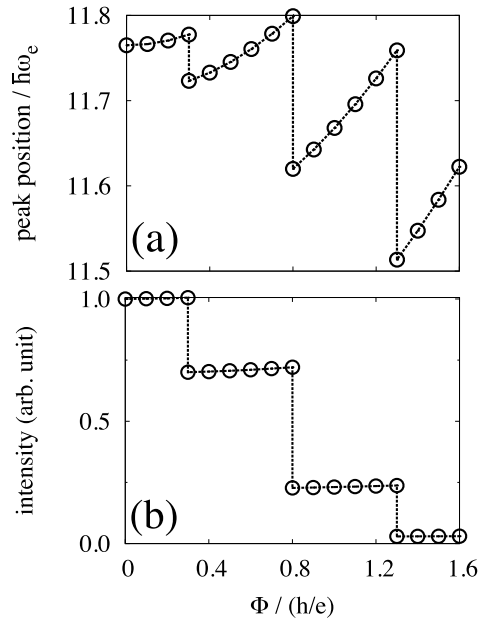}
  \caption{
    (a) The peak position and (b) intensity of the
    photoluminescence from a biexciton,
    as a function of the magnetic flux $\Phi$.
    The radius $R$, at which $V_\text{e}(r)$ takes a minimum,
    equals the effective Bohr radius $a_\text{\tiny B}$.
    $\Phi= \pi R^2 B$.
    \label{fig:biexciton}}
\end{figure}

Finally, we present the photoluminescence from a biexciton
in Fig.\ \ref{fig:biexciton}. In this case, the final state
is an excitonic state of the same angular momentum as the
ground state of the biexciton. The $\Phi$ dependence of
the peak position and intensity is qualitatively the same as
that for the photoluminescence from a trion. 

\section{CONCLUSIONS \label{sec:conclusions}}

We have examined the magnetoluminescence from a
trion and a biexciton in a type-II semiconductor quantum dot,
in which holes are confined inside the quantum dot and electrons
are in a ring-shaped region surrounding the quantum dot.
First, we have calculated the many-body states by the exact
diagonalization method. We have shown that the
two electrons in trions and biexcitons
form a Wigner molecule, reflecting a large correlation effect.
The electrons behave as a composite particle whose mass and charge
are twice those of a single electron. In consequence, the
ground-state energy of the trion and biexciton oscillates
as a function of magnetic flux $\Phi$ with a period of
approximately $h/(2e)$.
Next, we have evaluated the photoluminescence from the electron-hole
complexes as a function of $\Phi$.
Both the peak position and peak intensity of the
photoluminescence change discontinuously at the transition of the
ground state.
This indicates a possible observation of Wigner molecules by
optical experiment.

\begin{acknowledgments}
The authors acknowledge fruitful discussion with K.\ M.\ Itoh,
S.\ Miyamoto, and T.\ Ishikawa.
This work was partly supported by a Grant-in-Aid for Scientific
Research from the Japan Society for the Promotion of Science,
and Graduate School Doctoral Student Aid Program from Keio University.
\end{acknowledgments}

\appendix
\section*{APPENDIX: APPROXIMATE WAVE FUNCTIONS OF TWO ELECTRONS IN RING}

In a trion and a biexciton, two electrons are strongly correlated
to each other, 
forming a Wigner molecule.
In this appendix, we present a
simple wave function to describe the two-electron state,
neglecting the hole state. Using the wave function, we derive an
intensity ratio of the photoluminescence shown in Figs.\ 
\ref{fig:trion}(b) and \ref{fig:biexciton}(b).

We construct a wave function by a few electronic configurations
in which electrons occupy the lowest states in the radial motion,
$\psi_{\text{e},l,1}\chi_{\sigma}$.
In general, $\psi_{\text{e},l,n}$ is written as
\begin{equation} \tag{A1}
\psi_{\text{e},l,n} (\bm{r}) = R_{\text{e},l,n}(r) e^{i l \theta},
\end{equation}
where $R_{\text{e},l,n}$ is determined by the equation
\begin{widetext}
\begin{equation} \tag{A2}
\left\{ \frac{\hbar^2}{2 m_\text{e}} \left[
-\frac{\partial^2}{\partial r^2}
-\frac{1}{r}\frac{\partial}{\partial r}
+\left( \frac{l}{r} - \frac{eBr}{2\hbar} \right)^2
\right] + V_\text{e}(r)
\right\} R_{\text{e},l,n}
= \varepsilon_{\text{e},l,n} R_{\text{e},l,n},
\end{equation}
with $\varepsilon_{\text{e},l,n}$ being the energy eigenvalue for
the one-electron state.
Since the electron is confined around $r=R$ by $V_\text{e}(r)$,
both the centrifugal potential,
$\hbar^2 l^2/(2 m_\text{e} r^2)$, and diamagnetic term,
$(eBr)^2/(8m_\text{e})$, are ineffective.
Hence, (i) we replace $R_{\text{e},l,1}$ by $R_{\text{e},1}$,
disregarding its $l$ dependence hereafter.
(ii) The $B$ dependence of $R_{\text{e},1}$ is small. This
explains the plateau structure of photoluminescence intensity,
shown in Figs.\ \ref{fig:trion}(b) and \ref{fig:biexciton}(b),
in which the intensity is almost constant as long as the angular
momentum $l$ does not change.

Let us begin with the lowest state with the total angular
momentum $L=0$. Without the Coulomb interaction,
the lowest state is
$\hat e^\dagger_{0, +} \hat e^\dagger_{0, -} |0\rangle$,
where $\hat e^\dagger_{l,\sigma}$ is the creation operator of
state $\psi_{\text{e},l,1} \chi_{\sigma}$. The total spin is a
singlet. The wave function becomes
  \[
    \langle \bm{r}_1, \bm{r}_2 |
    \hat e^\dagger_{0, +} \hat e^\dagger_{0, -} |0\rangle
    = R_{\text{e}, 1}(r_1)R_{\text{e}, 1}(r_2)
    \frac{\chi_+(1) \chi_-(2) - \chi_-(1) \chi_+(2)}{\sqrt 2}.
  \]
This has a finite value at $\theta_1 = \theta_2$ since
no correlation effect is taken into account. We mix the
second lowest state with $L=0$,
$( \hat e^\dagger_{1,+}   \hat e^\dagger_{-1,-}
   - \hat e^\dagger_{1,-} \hat e^\dagger_{-1,+} )
|0\rangle / \sqrt2$,
as
  \begin{equation} \tag{A3}
    |L=0\rangle =
    \left[
      \sqrt{\frac{2}{3}} \hat e^\dagger_{0, +}
                         \hat e^\dagger_{0, -}
      - \sqrt{\frac{1}{6}} \left(
          \hat e^\dagger_{1,+}   \hat e^\dagger_{-1,-}
        - \hat e^\dagger_{1,-} \hat e^\dagger_{-1,+}
      \right)
    \right] |0\rangle.
  \label{eq:apdx-state0}
  \end{equation}
Then its wave function is given by
  \[
    \langle \bm{r}_1, \bm{r}_2 | L=0 \rangle
    = R_{\text{e}, 1}(r_1)R_{\text{e}, 1}(r_2)
    \frac{1-\cos(\theta_1-\theta_2)}{\sqrt{3/2}}
    \frac{\chi_+(1) \chi_-(2) - \chi_-(1) \chi_+(2)}{\sqrt 2},
  \]
\end{widetext}
which vanishes at $\theta_1 = \theta_2$. Thus this is an
appropriate state to describe the strongly correlated electrons
in the Wigner molecule.

We proceed to the lowest state with $L=1$.
In the absence of Coulomb interaction, it is
  \begin{equation} \tag{A4}
    |L=1\rangle
    = \frac{1}{\sqrt 2} \left(
          \hat e^\dagger_{0,+} \hat e^\dagger_{1,-}
          + \hat e^\dagger_{0,-} \hat e^\dagger_{1,+} \right)
          |0\rangle.
  \label{eq:apdx-state1}
  \end{equation}
The total spin is a triplet. The orbital part of the
wave function is
\[
R_{\text{e}, 1}(r_1) R_{\text{e}, 1}(r_2)
\frac{e^{i\theta_1} - e^{i\theta_2}}{2},
\]
the amplitude of which is zero at $\theta_1 = \theta_2$.
In the spin-triplet states, the exchange correlation
reduces the Coulomb energy between electrons.
Hence we adopt the state in Eq.\ (\ref{eq:apdx-state1})
as an approximate state for the Wigner molecule.

For the state with $L=2$, we mix two electronic configurations,
$\hat e^\dagger_{1, +} \hat e^\dagger_{1, -} |0\rangle$ and
$(\hat e^\dagger_{2,-}   \hat e^\dagger_{0,-}
- \hat e^\dagger_{2,+} \hat e^\dagger_{0,+})
|0\rangle / \sqrt2$, in such a way that the wave function
vanishes at $\theta_1 = \theta_2$. We obtain
  \begin{equation} \tag{A5}
    |L=2\rangle =
    \left[
      \sqrt{\frac{2}{3}} \hat e^\dagger_{1, +}
                         \hat e^\dagger_{1, -}
    - \sqrt{\frac{1}{6}} \left(
      \hat e^\dagger_{2,-} \hat e^\dagger_{0,-}
    - \hat e^\dagger_{2,+} \hat e^\dagger_{0,+}
    \right) \right] |0\rangle.
  \label{eq:apdx-state2}
  \end{equation}

The intensity of the photoluminescence from the states in Eqs.\
(\ref{eq:apdx-state0})--(\ref{eq:apdx-state2}) is evaluated
using Eq.\ (\ref{eq:intensity-formula}). The hole state is
$\hat h^\dagger_{0,+} |0\rangle$ or
$\hat h^\dagger_{0,-} |0\rangle$ in trion and
$\hat h^\dagger_{0,+} \hat h^\dagger_{0,-} |0\rangle$ in biexciton,
where $\hat h^\dagger_{0,\sigma}$ is the creation operator of
$\psi_{\text{h},0,1} \chi_{-\sigma} \equiv
R_{\text{h},1}(r) \chi_{-\sigma}$.
From $|L=0\rangle$, $|L=1\rangle$, and $|L=2\rangle$,
the intensity is given by $(2/3) I_0$, $(1/2) I_0$, and $(1/6) I_0$,
respectively, for the trion (they are twice for the biexciton).
Here,
\begin{equation} \tag{A6}
    I_0 = \frac{4}{3}
    \frac{E_\text{gap}^3}
    {4\pi\epsilon_0 \hbar^4 c^3}
    \left| d_\text{vc} \int 2 \pi r \mathrm{d}r ~ R_{\text{e},1}(r)
    R_{\text{h},1}(r) \right|^2,
  \end{equation}
where $E_\text{gap}$ is the band gap and 
$d_\text{vc}$ is given by Eq.\ (\ref{eq:d-vc}).

The lowest states with $L>2$ do not include the one-electron state with
$l=0$ ($\psi_{\text{e},0,1}\chi_{\sigma}$) in our approximation.
As a result, the intensity of the photoluminescence becomes zero.
In conclusion, the ratio of the intensity is
  \begin{equation} \tag{A7}
    I(L=0):I(L=1):I(L=2):I(L>2)
    = ~4~:~3~:~1~:~0.
  \end{equation}


\end{document}